# Ultimate low system dark count rate for superconducting nanowire single-photon detector

Hiroyuki Shibata,[1,*] Kaoru Shimizu,[2] Hiroki Takesue,[2] and Yasuhiro Tokura[2]

[1]Dept. of Electrical and Electronic Engineering, Kitami Institute of Technology, 165 Koen-cho, Kitami, Hokkaido 090-8507, Japan
[2]NTT Basic Research Laboratories, NTT Corporation, 3-1 Morinosato-Wakamiya, Atsugi, Kanagawa 243-0198, Japan
*Corresponding author: shibathr@mail.kitami-it.ac.jp

The dark count rate (DCR) is a key parameter of single-photon detectors. By introducing a bulk optical band-pass filter mounted on a fiber-to-fiber optical bench cooled at 3 K and blocking down to 5 μm, we suppressed the DCR of a superconducting nanowire single-photon detector by more than three orders of magnitude. The DCR is limited by the blackbody radiation through a signal passband of 20 nm bandwidth. The figure of merit, system detection efficiency, and DCR were $2.7 \times 10^{11}$, 2.3%, and $10^{-3}$ Hz, respectively. Narrowing the bandwidth to 100 GHz suppresses the DCR to $10^{-4}$ Hz and the figure of merit increases to $1.8 \times 10^{12}$.

Superconducting single-photon detectors (SNSPDs or SSPDs) based on ultrathin superconducting nanowires are now widely recognized as high-performance single-photon detectors in the near-infrared region. The high detection efficiency ($\eta$), low dark count rate (DCR), wide detection wavelength, and small timing jitter ($\Delta t$) of SSPDs are now invaluable for many fields such as quantum information, quantum optics, free space laser communication, and biological applications [1–10].

One of the key parameters of the SNSPD is the DCR. The system DCR is the sum of an intrinsic DCR and $DCR_{black}$ [11–14]. Here, the system DCR is the response pulse count rate when the input fiber to the cryocooler is blocked, the intrinsic DCR is the response pulse count rate when the fiber to the device is disconnected inside the cryocooler, and $DCR_{black}$ is the pulse count rate due to the blackbody radiation at room temperature that propagates through the optical fiber to the device. The intrinsic DCR increases exponentially when the bias current approaches the critical current but is below 1 Hz in the low bias region. On the other hand, $DCR_{black}$ is around 10 to 1 kHz owing to the wide detection wavelength and high sensitivity of SNSPDs. Thus, the system DCR is usually dominated by $DCR_{black}$ in the low bias region. Previously, we successfully suppressed the system DCR by two orders of magnitude using a cold pigtailed band-pass filter to block $DCR_{black}$ and applied this result to a long-distance quantum key distribution experiment over 72 dB channel loss [7,11]. Then, Yang et al. also suppressed the system DCR by two orders of magnitude by applying an on-chip band-pass filter to the SNSPD device [12,13]. However, owing to the imperfect blocking properties of the filters, the system DCR was still limited by $DCR_{black}$ at wavelengths longer than 2 μm [11–17]. To further reduce the system DCR, it is desirable to suppress $DCR_{black}$ completely.

In this paper, we report an SNSPD with the ultimate low DCR, which is obtained by completely blocking the blackbody radiation except in the signal passband using a bulk band-pass filter mounted on a fiber-to-fiber optical bench cooled at 3 K. We also study the effect of reducing the passband on the system DCR.

Figure 1(a) shows the transmission spectra of the bulk band-pass filter (Andover, 200FC40) at room temperature obtained using a Fourier transform infrared spectrometer (Bruker, IFS-113V). It has a transmission window of 20 nm at 1550-nm and a rejection band down to 5000 nm. The bulk filter is mounted on a fiber-to-fiber bench made of copper and fixed at the second stage of the [3]He cryocooler. We put two bulk filters in series to enhance the rejection properties. Figure 1(b) shows the transmission spectra of the band-pass filters obtained using a standard optical spectrum analyzer (Agilent 70951A). It has a bandwidth of 20 nm at a 1550 nm wavelength and an insertion loss of 2.3 dB at room temperature, including the coupling loss at the cryocooler. As the temperature decreases to 3 K, the insertion loss increases to 3.9 dB and the passband shifts 8 nm lower. Since light longer than 5000 nm in wavelength cannot be transmitted in standard telecom fiber, we can reject $DCR_{black}$ completely using the cold band-pass filters except for the signal passband.

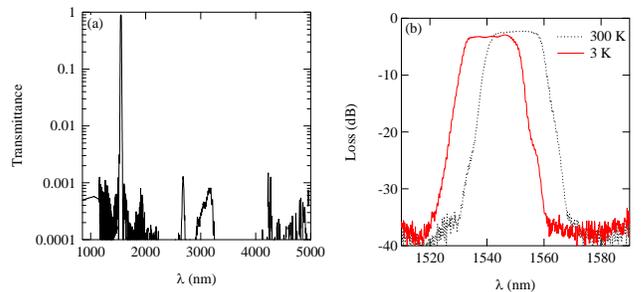

Fig. 1. (a) FT-IR transmission spectra of the bulk band-pass filter in a wide wavelength region at room temperature. (b) Transmission spectra of the filters mounted on a fiber-to-fiber bench at room temperature (black dots) and at 3 K (red line) obtained using a standard optical spectrum analyzer.

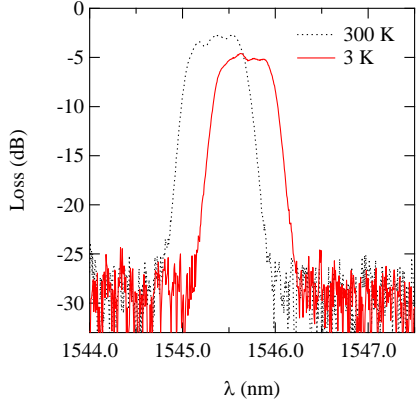

Fig. 2. Transmission spectra of DWDM filter in series with bulk filters mounted on a fiber-to-fiber bench at room temperature (black dots) and at 3 K (red line) using a standard optical spectrum analyzer.

The bandwidth of the filter can be further reduced by adding a cold dense wavelength division multiplexer (DWDM) filter with a 100 GHz bandwidth (Advanced Fiber Resources, DWDM-1-40). Figure 2 shows the transmission spectra of the DWDM filter in series with the bulk filters. At room temperature, the bandwidth is reduced to 100 GHz at a wavelength of 1545.3 nm, and the insertion loss increases to 3.0 dB. The loss increases to 4.5 dB as the temperature decreases to 3 K, and the passband shifts 0.3 nm higher. A slight increase in the passband during cooling has also been observed for a coarse wavelength division multiplexer filter [11].

The SNSPD device is fabricated as follows. NbN thin films 7 nm in thickness were synthesized on a thermally oxidized Si substrate by reactive DC magnetron sputtering. The $SiO_2$ layers on both sides of the Si substrate are about 260 nm thick and serve as an antireflection layer and a front-side cavity structure at telecom wavelengths, respectively. The NbN meander pattern was fabricated with a standard e-beam process using a negative resist (SNR-M5) and $C_2F_5$ dry etching. The size of the meander is $15 \times 15$ μm$^2$ with a line and space width of 100 nm. A backside cavity structure was also formed on the meander using a standard photolithography process to enhance the detection efficiency. The backside cavity is composed of a 260-nm-thick $SiO_2$ layer, 2-nm-thick Ti, and 100-nm-thick Au.

For optical characterization, we used a $^3$He cryocooler with a base temperature of 0.38 K. A continuous-wave tunable laser was used to measure $\eta$ and the DCR. The power was strongly attenuated to $10^6$ photons/s and was focused on the device from the substrate side of the device using a lensed optical fiber. To focus the beam on the meander pattern, we observed the optical spot using an infrared microscope. The wavelength was fixed at 1545.6 nm, which was the highest transmission of the filter at 3 K. The output signal from the SNSPD was amplified by two room-temperature rf amplifiers with a total gain of 60 dB and sent to a gated photon counter. To measure the jitter, we used a femtosecond fiber laser with a repetition rate of 100 MHz and a time interval analyzer. Details of our jitter measurement system were published earlier [18].

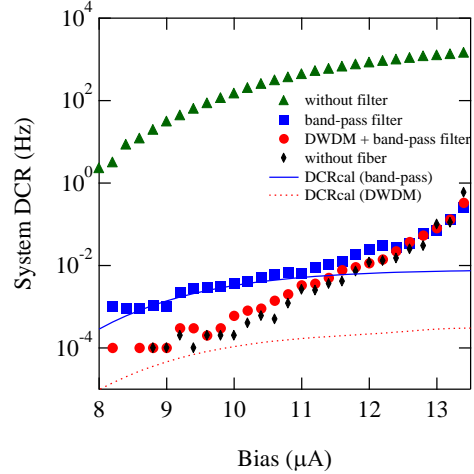

Fig. 4. Bias current dependence of system DCR without filters (green triangles), with band-pass filter mounted on a fiber-to-fiber bench at 3 K (blue squares), with both band-pass filter and DWDM cold filter (red circles), and without connecting fiber, indicating intrinsic DCR (black diamonds). Thin blue line is the calculated DCR due to the blackbody radiation passing through the band-pass filter with 20 nm bandwidth; thin red dotted line is the calculated DCR due to the blackbody radiation passing through both the band-pass filter and DWDM filter with 100 GHz bandwidth

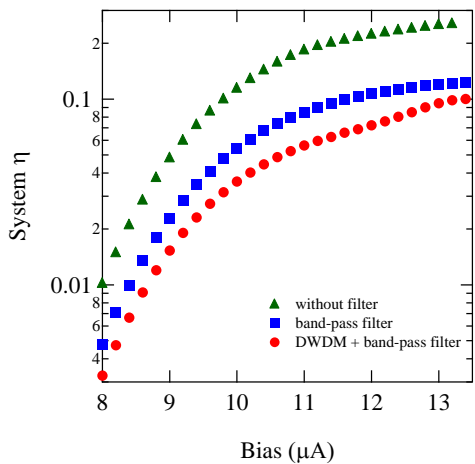

Fig. 3. Bias current dependence of system detection efficiency ($\eta$) without filters (green triangles), with band-pass filter mounted on a fiber-to-fiber bench at 3 K (blue squares), and with both band-pass filter and DWDM cold filter (red circles).

Figure 3 shows the bias current dependence of the system $\eta$ of the device. The system $\eta$ reaches 26% as the bias current increases to 13.2 μA. The system $\eta$ is not high compared to that of a state-of-the-art SNSPD with a double-side cavity structure [9,10]. This may be due to the greater cross-section of our nanowires and less optimized

SiO$_2$ thickness. As the bulk band-pass filter cooled to 3 K is introduced before the device, the system $\eta$ decreases to 12% owing to the loss of the filter (blue squares) and decreases to 10% as the DWDM filter is added (red circles).

The bias current dependence of the system DCR is shown in Fig. 4. The system DCR decreases by more than three orders of magnitude as the band-pass filter is introduced (blue squares). The value is more than one order of magnitude lower than that of our previous report using a band-pass filter rejecting down to 2 μm and confirms the effect of blackbody radiation below 2 μm [11]. The black diamonds show the results for the DCR without the connecting fiber, indicating the intrinsic DCR of the SNSPD. The thin blue line is the calculated DCR due to the blackbody radiation passing through the band-pass filter with a bandwidth of 20 nm [11]. It is clear that the system DCR with the band-pass filter is dominated by the intrinsic DCR in the high bias region (>12 μA) and is dominated by the DCR due to the blackbody radiation passing through the passband in the low bias region (<11 μA). In Fig. 4, the bias dependence of the system DCR with both the DWDM filter and band-pass filter is also shown (red circles). It follows almost the same line as that without the connecting fiber, indicating that the system DCR with both the DWDM and band-pass filter is determined by the intrinsic DCR. Actually, the calculated DCR$_{black}$ with the 100 GHz passband (thin red line) is lower than the intrinsic DCR for all the bias regions. In the low bias region (<11 μA), the system DCR decreases by more than four orders of magnitude compared to that without filters, and it is reduced to 10$^{-4}$ Hz at a bias current of 9 μA.

The bias current dependence of the timing jitter ($\Delta t$) is shown in Fig. 5. The value is about 50 to 100 ps, and it decreases as the bias current increases. These features are almost the same as those previously reported [9]. In the figure, the timing jitter increases by about 10 ps as the filters are introduced. This may be due to the distortion of the pulse shape. The bandwidth of the femtosecond pulse

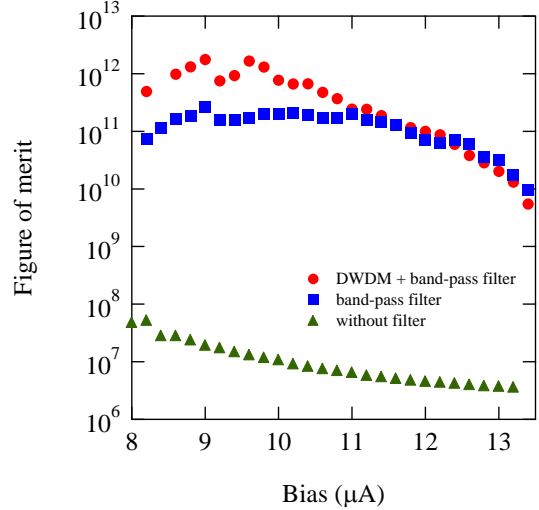

Fig. 6. Bias current dependence of the figure of merit [$\eta/(DCR\Delta t)$] without filters (green triangles), with the cold band-pass filter (blue squares), and with both the cold band-pass filter and cold DWDM filter (red circles).

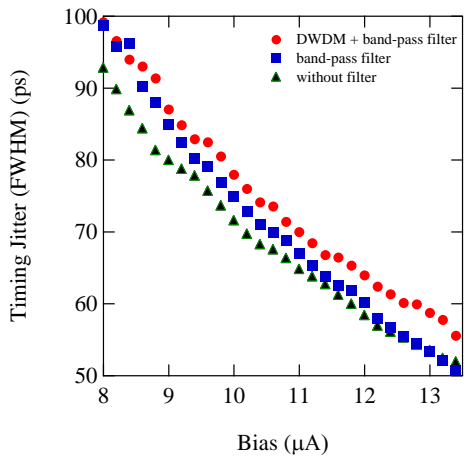

Fig. 5. Bias current dependence of the timing jitter ($\Delta t$) without filters (green triangles), with the cold band-pass filter (blue squares), and with both the cold band-pass filter and cold DWDM filter (red circles).

laser is about 200 nm, which is much larger than the passband of the cold filters.

The figure of merit of a single-photon detector is known to be well represented as $\eta/(DCR\Delta t)$ [3,4]. Figure 6 summarizes the bias current dependence of the figure of merit. Introducing the cold band-pass filter increases it by more than three orders even in the high bias region (>12 μA). In the low bias region (<11 μA), it is almost constant, as the system DCR is dominated by DCR$_{black}$. It reaches $2.7 \times 10^{11}$ with a system $\eta$ of 2.3% and system DCR of $10^{-3}$ Hz at a bias current of 9 μA. This is the highest reported value as far as we know. Adding the DWDM filter increases it further in the low bias region (<11 μA) owing to a further decrease in the system DCR, and it reaches $1.8 \times 10^{12}$ with a system $\eta$ of 1.5% and system DCR of $10^{-4}$ Hz.

There are two ways to further increase the figure of merit of the SNSPD. One is to increase the $\eta$ value by refining the fabrication process. The other is to reduce the loss of the cold optical filters. This can be achieved by designing and fabricating a low-loss filter with the appropriate rejection properties and passband on the device instead of introducing commercial filters.

In conclusion, we demonstrate that introducing a cold optical filter with an appropriate passband is quite effective for reducing the system DCR of an SNSPD. By completely blocking the blackbody radiation except in the signal passband of 20 nm, the system DCR can be suppressed by more than three orders of magnitude. An SNSPD with a figure of merit of $2.7 \times 10^{11}$, $\eta = 2.3\%$, and DCR = $10^{-3}$ Hz is obtained. By narrowing the passband to 100 GHz using a cold DWDM filter, an SNSPD with a figure of merit of $1.8 \times 10^{12}$, $\eta = 1.5\%$, and DCR = $10^{-4}$ Hz is also obtained.